\documentclass[aps,prl,showpacs,floatfix,
               amsmath,amssymb,amsfonts,nobalancelastpage,superscriptaddress,
			   tightenlines,reprint]{revtex4-1}
\usepackage{graphicx}
\usepackage{color}
\usepackage{ulem}
\usepackage{bm}
\usepackage[colorlinks=true]{hyperref}
\usepackage{latexsym}
\usepackage{dsfont}
\usepackage{epstopdf}
\usepackage{CJK}

\newcommand{\bfs}{BaFe$_2$Se$_3$}
\newcommand{\mytitle}{Block magnetic excitations in the orbitally-selective Mott insulator BaFe$_2$Se$_3$}
\newcommand{\be}{\begin{equation}}
\newcommand{\ee}{\end{equation}}
\newcommand{\bea}{\begin{eqnarray}}
\newcommand{\eea}{\end{eqnarray}}

\def\Ang{\AA$^{-1}$}

\begin{document}
%---------------------------------------------------------------------
% Preamble
%---------------------------------------------------------------------
\title{\mytitle}
% ------------------
\author{M. Mourigal}
\altaffiliation{\textit{Present address:} 
             School of Physics, Georgia Institute of Technology, Atlanta, GA 30332, USA}
\affiliation{Institute for Quantum Matter 
             and Department of Physics and Astronomy,
             The Johns Hopkins University, Baltimore, MD 21218, USA}    
% ------------------
\author{Shan Wu}
\affiliation{Institute for Quantum Matter 
             and Department of Physics and Astronomy,
             The Johns Hopkins University, Baltimore, MD 21218, USA}	
% ------------------
\author{M. B. Stone}
\affiliation{Quantum Condensed Matter Division, Neutron Sciences Directorate, 
			 Oak Ridge National Laboratory, Oak Ridge, TN 37831, USA}               
% ------------------
\author{J. R. Neilson}
\altaffiliation{\textit{Present address:}  Department of Chemistry, Colorado State University, Fort Collins, CO 80523, USA}
\affiliation{Institute for Quantum Matter 
             and Department of Physics and Astronomy,
             The Johns Hopkins University, Baltimore, MD 21218, USA}
\affiliation{Department of Chemistry,
             The Johns Hopkins University, Baltimore, MD 21218, USA}              
% ------------------
\author{J. M. Caron}
\altaffiliation{\textit{Present address:} Department of Materials Science and Engineering, 
Cornell University, Ithaca, NY 14853, USA}
\affiliation{Institute for Quantum Matter 
             and Department of Physics and Astronomy,
             The Johns Hopkins University, Baltimore, MD 21218, USA}
\affiliation{Department of Chemistry,
             The Johns Hopkins University, Baltimore, MD 21218, USA}
% ------------------             
\author{T. M. McQueen}
\affiliation{Institute for Quantum Matter 
             and Department of Physics and Astronomy,
             The Johns Hopkins University, Baltimore, MD 21218, USA} 
\affiliation{Department of Chemistry,
             The Johns Hopkins University, Baltimore, MD 21218, USA}	                
\affiliation{Department of Materials Science and Engineering,
             The Johns Hopkins University, Baltimore, MD 21218, USA}  
% ------------------
\author{C. L. Broholm}
\affiliation{Institute for Quantum Matter 
             and Department of Physics and Astronomy,
             The Johns Hopkins University, Baltimore, MD 21218, USA}  
\affiliation{Quantum Condensed Matter Division, Neutron Sciences Directorate, 
		     Oak Ridge National Laboratory, Oak Ridge, TN 37831, USA}
\affiliation{Department of Materials Science and Engineering,
             The Johns Hopkins University, Baltimore, MD 21218, USA} 		     
% --------------------------------------------------------------------						
\date{\today}
% --------------------------------------------------------------------
\begin{abstract} 
      {Iron pnictides and selenides display a variety of unusual magnetic phases originating from the interplay between electronic, orbital and lattice degrees of freedom. Using powder inelastic neutron scattering on the two-leg ladder BaFe$_2$Se$_3$, we fully characterize the static and dynamic spin correlations associated with the Fe$_4$ block state, an exotic magnetic ground-state observed in this {low-dimensional} magnet and in Rb$_{0.89}$Fe$_{1.58}$Se$_{2}$. All the magnetic excitations of the Fe$_4$ block state predicted by an effective Heisenberg model with localized spins are observed below 300 meV and quantitatively reproduced. However, the data only accounts for 16(3) $\mu_{\rm B}^2$ per Fe$^{2+}$, approximatively $2/3$ of the total spectral weight expected for localized $S=2$ moments. Our results highlight how orbital degrees of freedom in iron-based magnets can conspire to stabilize an exotic magnetic state.}
\end{abstract}
% --------------------------------------------------------------------
% --------------------------------------------------------------------
\maketitle
% --------------------------------------------------------------------

%---------------------------------------------------------------------
% Main Text
%---------------------------------------------------------------------

Magnetism in iron-based superconductors is a complex many-body phenomenon and understanding it is now a central challenge in condensed-matter physics~\cite{Lumsden2010a,Dai2012a}. The parent compounds of a large majority of iron-based superconductors are constructed from quasi two-dimensional (2D) FeAs or FeSe layers and host a range of metallic, semi-metallic, and insulating behaviors originating from the interplay between structural, orbital, magnetic and electronic degrees of freedom~\cite{Chubukov2008a,Lee2009a,Fernandes2012a,Dagotto2013a}. Their magnetic ground-states and excitations have been extensively studied by neutron scattering~\cite{Ewings2008a,Zhao2009a,Harriger2011a,Lipscombe2011a,Wang2011a,Liu2012a} but a unified theoretical description that accounts for the role of Coulomb repulsion and Hund's coupling on electrons occupying multiple active $3d$-orbitals remains a challenging task. To understand these unfamiliar Fe-based magnets it is necessary to probe electronic correlations at the atomic scale in chemically and structurally related materials.

Inspired by the successful description of magnetic, electronic and orbital phenomena in various insulating chain- and ladder-based cuprates~\cite{Abbamonte2004a,Lake2010a,Schlappa2012a}, recent experimental work explored the properties of {structuraly} quasi one-dimensional (1D) Fe-based compounds such as  $\rm KFe_2Se_3$~\cite{Caron2012a}, $\rm CsFe_2Se_3$~\cite{Du2012a}, $\rm BaFe_2Se_2O$~\cite{Popoviic2014a}, $\rm TaFe_{1+y}Te_3$~\cite{Ke2012a} and the two-leg ladder \bfs~\cite{Caron2011a,Nambu2012a,Saparov2011a,Lei2011a,Lei2012a,Monney2013a}. Unlike the former materials, \bfs\ hosts an exotic form of magnetic order, the Fe$_4$ \textit{block state}, that has also been observed in the $\sqrt{5}\!\times\!\sqrt{5}$ vacancy-ordered quasi-2D compound $\rm Rb_{0.89}Fe_{1.58}Se_{2}$~\cite{Ye2011a,Wang2011a} and reproduced by first-principles electronic structure calculations~\cite{Yin2012a}. Facilitated by low-dimensionality, exact diagonalization (ED) and density-matrix renormalization group (DMRG) analysis~\cite{Luo2013a,Rincon2014a} of multi-orbital Hubbard models relevant for \bfs\ indicate the exotic block state is stabilized by sizable Hund's coupling~\cite{Luo2013a,Rincon2014a}. It is proposed that \bfs\ forms an \textit{orbital-selective Mott phase}~\cite{Rincon2014a} where narrow-band localized electrons coexist with wide-band itinerant electrons originating from different $3d$ atomic orbitals~\cite{Yin2012a,Georges2013a}. 

% --------------------------------------------------------------------
\begin{figure}[t!]
\includegraphics[width=0.90\columnwidth]{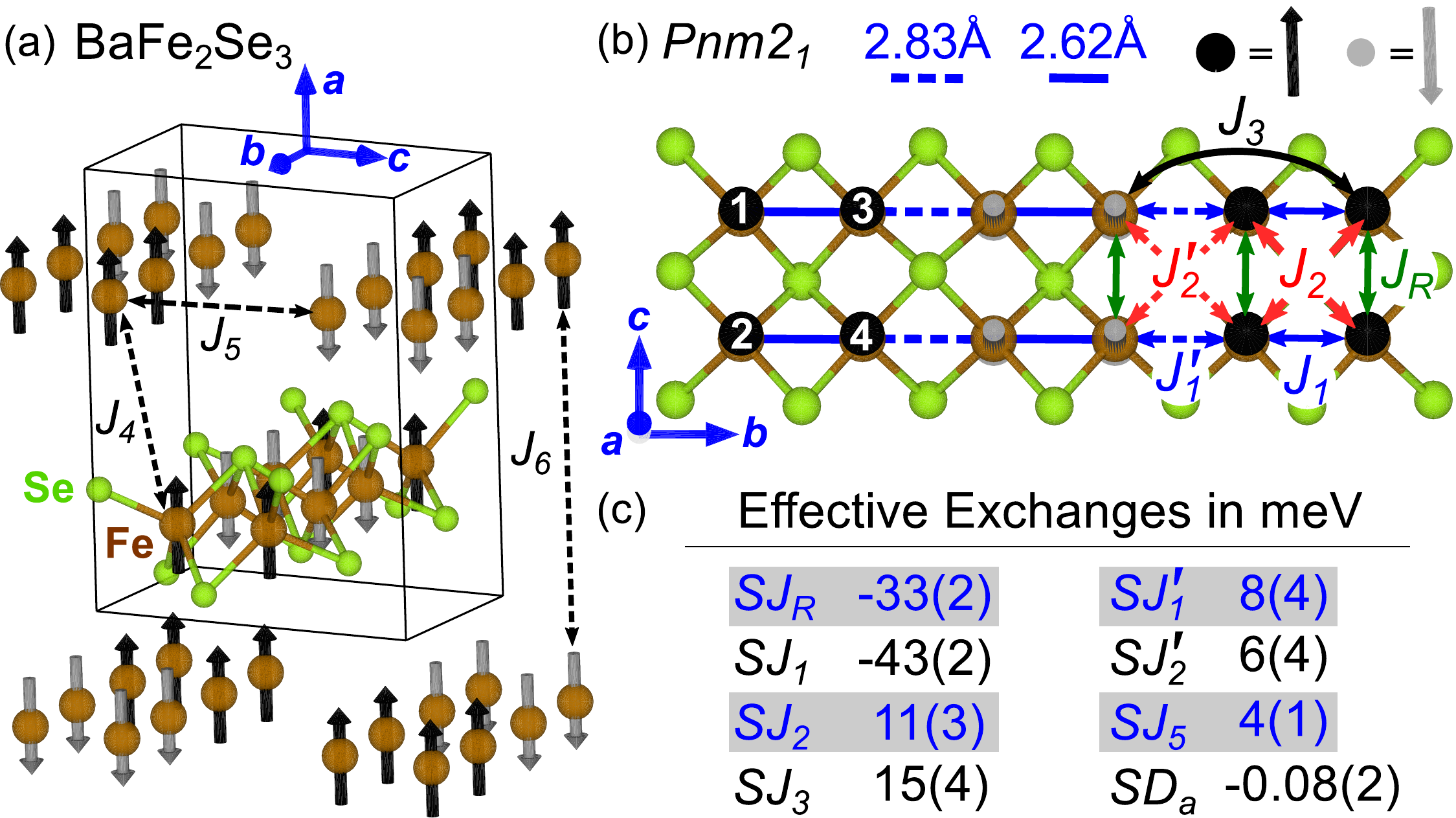}
\caption{(Color online) (a) Crystal structure of \bfs\ with $a\!=\!11.88$~\AA, $b\!=\!5.41$~\AA\ and $c\!=\!9.14$~\AA. Ba-atoms are omitted. The Fe$_4$ block ground-state is represented with light (spin-down) and dark (spin-up) bold arrows. (b) Structure of an individual ladder. (c) Values of exchange interactions determined from our data using an effective Heisenberg model.}\vspace{-0.6cm}
\label{fig1}
\end{figure} 
% --------------------------------------------------------------------

In this work, we determine the magnetic excitation spectrum of \bfs\ through broad band inelastic neutron scattering from a powder sample. We provide direct spectroscopic evidence for the Fe$_4$ block state [Fig.~\ref{fig1}] and develop an effective Heisenberg model that accounts for all observed acoustic and optical spin-wave modes. We also determine the effective moment in the energy range below $300$~meV to be $\mu_{\rm eff}^2\approx16$~$\mu^2_{\rm B}$ per Fe, which is indicative of spin-orbital magnetism in \bfs.  

The crystal structure of \bfs\ [Fig.~\ref{fig1}] determined by low-temperature neutron powder diffraction (NPD)~\cite{Krzton-Maziopa2011a,Lei2011a,Caron2011a,Nambu2012a} comprises edge-sharing FeSe$_4$ tetrahedra forming two-leg Fe-ladders organized in a face-centered orthorhombic lattice. Neutron pair-distribution-function (NPDF) analysis~\cite{Caron2011a,Caron2012a} reveals gradual Fe displacements upon cooling leading to \textit{structural} Fe$_4$ blocks (plaquettes)~[Fig.~\ref{fig1}(b)] with two inequivalent Fe sites in the $Pnm2_1$ space-group. The Fe environments are distorted with four distinct distances to coordinating Se atoms. \bfs\ is an insulator with a resistivity activation-energy of $E_a\!\approx\!0.13$--0.18~eV~\cite{Lei2011a,Nambu2012a}. Assuming a high-spin electronic configuration in the tetrahedral crystal field leads to $S\!=\!2$ per Fe$^{2+}$ $3d^6$ ions. Long-range magnetic order develops below $T_{\rm N}\!\approx\!255$~K with a saturated moment of $2.8$~$\mu_{\rm B}$ per Fe~\cite{Caron2011a,Nambu2012a} and a propagation-vector $\boldsymbol{\kappa}=(\frac{1}{2},\frac{1}{2},\frac{1}{2})$. The corresponding magnetic structure consists of Fe$_4$ blocks in which four nearest-neighbor spins co-align parallel to the $\boldsymbol{a}$ direction~\cite{Caron2011a,Nambu2012a}. In turn, the plaquettes arrange in a staggered fashion along the ladder and inter-ladder directions [Fig.~\ref{fig1}(a)] with no net magnetization. This exotic magnetic state points to exchange frustration, orbital ordering and/or spin-lattice coupling and it was recently proposed that \bfs\ may host a ferrielectric polarization driven by exchange striction~\cite{Dong2014a}.

To search for magnetic excitations associated with the Fe$_4$ block spin structure, our inelastic neutron scattering experiment was performed on the ARCS~\cite{Stone2014a} time-of-flight spectrometer at the Spallation Neutron Source (SNS), Oak Ridge National Laboratory (ORNL). A $m\!\approx\!9.9$~g power sample of \bfs, synthesized using the method of Ref.~\cite{Caron2011a}, was mounted in a cylindrical aluminum can,  sealed under 1 atm of $^4$He, and cooled to $T\!=\!5$~K in a close-cycle cryostat. To reduce multiple scattering, the can contained horizontal sheets of neutron absorbing Cd inserted every centimeter between layers of \bfs\ powder. Data were acquired with the incident neutron energy set to $E_i\!=$ 20 meV, 50 meV, 150 meV and 450 meV with full-width at half-maximum (FWHM) elastic energy-resolution of 0.8 meV, 2.0 meV, 5.8 meV and 40 meV, respectively, and momentum-resolution of 0.060(4) \AA$^{-1}$, 0.079(6) \AA$^{-1}$, 0.11(1) \AA$^{-1}$, and 0.20(3) \AA$^{-1}$ estimated from the average FWHM of the four strongest nuclear reflections below $Q\leq2.2$~\AA$^{-1}$. The intensity measured as a function of momentum $\hbar Q=\hbar|{\bf Q}|$ and energy-transfer $E=\hbar\omega$, $\tilde{I}(Q,E)\!=\!k_i/k_f({\rm d}^2\sigma/{\rm d}\Omega {\rm d}E_f)$, was normalized to absolute units (mbarn sr$^{-1}$ meV$^{-1}$ Fe$^{-1}$) using the integrated intensity of nuclear Bragg scattering in the paramagnetic phase at $T=300$~K. This method was preferred over normalization to a Vanadium standard to ensure a reliable cross-calibration between datasets with very different $E_i$.

% --------------------------------------------------------------------
\begin{figure}[t!]
\includegraphics[width=0.95\columnwidth]{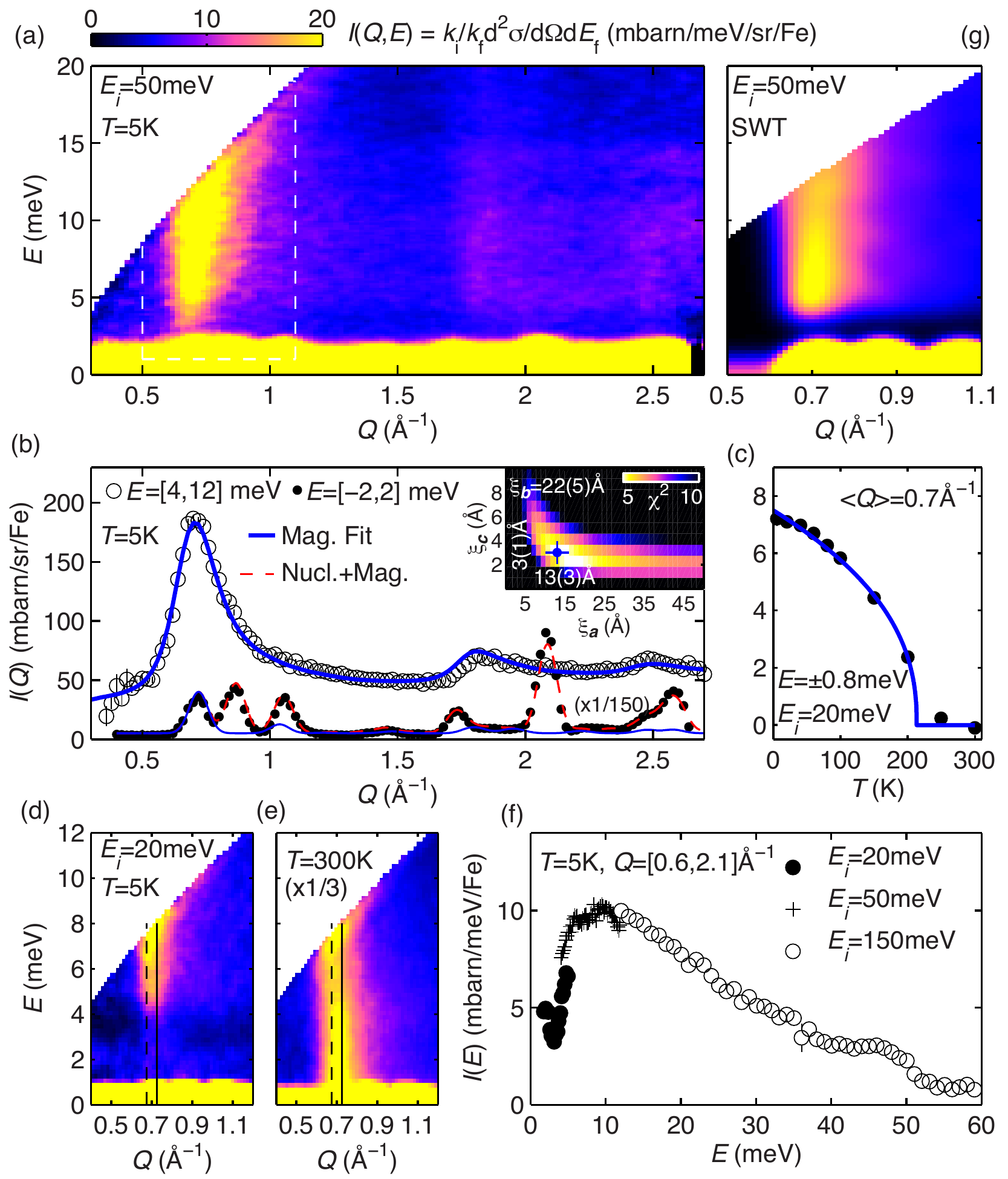}
\caption{(Color online) Low-energy spectrum of \bfs. (a) $\tilde{I}(Q,E)$ at $T\!=\!5$~K with $E_i\!=\!50$~meV. (b) $E$-integrated inelastic scattering $\tilde{I}({\it Q})$ (open symbols) and elastic scattering $\tilde{I}_0({\it Q})$ multiplied by a factor $1/150$ (full symbols) compared to model of Eq.~\ref{eq} {(solid-blue line)} and elastic scattering (dashed-red line). (inset) Fit {$\chi^2$ versus $\xi_a$ and $\xi_c$ and best fit values (blue symbol)}. (c) $T$-dependence of the integrated magnetic intensity measured with $E_i\!=\!20$~meV (full symbols) {and magnetic order parameter as a guide to the eye (blue line)}. (d-e) $\tilde{I}(Q,E)$ with $E_i\!=\!20$~meV at $T\!=\!5$~K and $T\!=\!300$~K. (f) Momentum-integrated inelastic scattering with different $E_i$'s. (g) SWT prediction $\tilde{I}_{\rm SWT}(Q,E)$ for $E_i\!=\!50$~meV in arbitrary intensity units.} \vspace{-0.8cm}
\label{fig2}
\end{figure} 
% --------------------------------------------------------------------

In Fig.~\ref{fig2}, we discuss the elastic scattering and spectrum of low-energy excitations measured in \bfs. At $T\!=\!5$~K [Fig.~\ref{fig2}(a)], we observe an intense ridge of inelastic signal which extends from $E\!\approx\!4$~meV and is characterized by a sharp onset at $Q\!\approx\!0.7$\Ang and a more gradual decay towards larger $Q$. A similar assymetric lineshape is observed for $Q\approx\!1.8$~\Ang\ and $2.5$~\Ang. The corresponding $\tilde{I}({\it Q})$ obtained by $E$-integration over the range $4\leq\!E\!\leq12$~meV is compared to the scaled elastic signal $\tilde{I}_{0}({\it Q})$ integrated over $E\!=\!\pm2$~meV in Fig.~\ref{fig2}(b). The coherent elastic signal can be reproduced without any free parameter using the known magnetic propagation vector $\boldsymbol{\kappa}$ and Fe$_4$ block spin structure, and a static moment of $\langle m\rangle\!=\!2.7(1)$~$\mu_{\rm B}$ per Fe. {The latter value is extracted from the low temperature limit for $|E|!<\!0.8$~meV [Fig.~\ref{fig2}(c)], corresponds to a static spin value $\langle S\rangle = 1.3(1)$ for $g=2$, and agrees remarkably well with the value obtained by neutron diffraction~\cite{Caron2011a,Nambu2012a,Caron2012a}}. As maxima in $\tilde{I}({\it Q})$ are observed around strong magnetic Bragg reflections of $\tilde{I}_{0}({\it Q})$, the former signal clearly originates from acoustic spin-waves and contains information particularly about inter-plaquette magnetic interactions in \bfs. 

To determine the dimensionality of magnetism in \bfs, we use an approach employed for spatially short-range ordered states~\cite{Warren1941a,Nakatsuji2005a} and compare the $E$-integrated inelastic signal to the {spherical average of the orthorhombic scattering cross-section}
\begin{equation}
	\nonumber
	\tilde{I}(Q)\!\propto\!\displaystyle{\int} \frac{{\rm d}\Omega}{4\pi} 
	|f({\bf Q})|^2 \displaystyle{\sum_{\boldsymbol{\tau}_m}}  
	\frac{|{\bf F}_{\!\perp\!}(\boldsymbol{\tau}_m)|^2}
	{\Big(1+\displaystyle{\sum_{\alpha=1}^3} \xi^2_\alpha 
	\left[ ({\bf Q} - {\boldsymbol{\tau}_m}) \cdot {\hat{\boldsymbol{e}}_\alpha} \right]^2\Big)^2 } 
	\ \mbox{,} \ \ 
	\label{eq}
\end{equation}
where $f({\bf Q})$ is the magnetic form-factor of Fe$^{2+}$, ${\bf F}_{\!\perp\!}({\bf Q})$ is the magnetic scattering amplitude perpendicular to the momentum-transfer ${\bf Q}$, and $\boldsymbol{\tau}_m=\boldsymbol{\tau}\pm\boldsymbol{\kappa}$ with $\boldsymbol{\tau}=h\boldsymbol{a}^*+k\boldsymbol{b}^*+\ell\boldsymbol{c}^*$ a reciprocal-lattice vector. The parameters $\xi_\alpha$ with $\alpha=1,2,3$ are pseudo-correlation lengths along the crystallographic directions $\boldsymbol{a}$, $\boldsymbol{b}$ and $\boldsymbol{c}$, respectively. {A fit to the asymmetric inelastic profile for $Q\!\leq\!2.8$~\AA\ [Fig.~\ref{fig2}(b)], for which the Ewald sphere passes through $\approx\!200$ magnetic peaks at distinct angles, robustly identifies anisotropic correlation lengths $\xi_{\boldsymbol{a}}=3(1)$~\AA, $\xi_{\boldsymbol{b}}=22(5)$~\AA\ and $\xi_{\boldsymbol{c}}=13(3)$~\AA\ [Fig.~\ref{fig2}(b)-inset]. In terms of inter-plaquette distances in the orthorhombic unit-cell [Fig.~\ref{fig1}], these correspond to $\xi_{\boldsymbol{b}}=4.1(8)~\boldsymbol{b}$, $\xi_{\boldsymbol{c}}=1.4(3)~\boldsymbol{c}$ and $\xi_{\boldsymbol{a}}=0.3(1)~\boldsymbol{a}$. Qualitatively, this reveals that \bfs\ is a low-dimensional magnet with a hierarchy of interactions that result in zero-dimensional Fe$_4$ blocks arranged as quasi-one-dimensional ladders extending along $\boldsymbol{b}$ that in turn interact weakly with their nearest neighbors to form a quasi-two-dimensional spin system in the $bc$-plane.}

Our higher-resolution $E_i\!=\!20$~meV data reveal an apparent gap $\Delta\!\approx\!5$~meV in the spectrum for $T\!=\!5$~K [Fig.~\ref{fig2}(d)]. While this gap closes when warming to $T\!=\!300$~K, the $\Delta$ energy scale remains apparent. The signal's lineshape changes from a Gaussian peak centered at $Q\approx0.72$~\Ang $=\!|\boldsymbol{\kappa}|$ for $E\!<\!\Delta$ to an asymmetric peak-shape that onsets at $\!Q\approx\!0.68$~\Ang\ $=\!|\left(0,\frac{1}{2},\frac{1}{2}\right)|$ for $E\!>\!\Delta$, where it resembles the lineshape of the low temperature spectrum. This behavior can be qualitatively understood as a consequence of the dimensionality of the inter-ladder interactions and a small single-ion or exchange anisotropy responsible for~$\Delta$. 

\begin{figure}[t!]
\includegraphics[width=0.95\columnwidth]{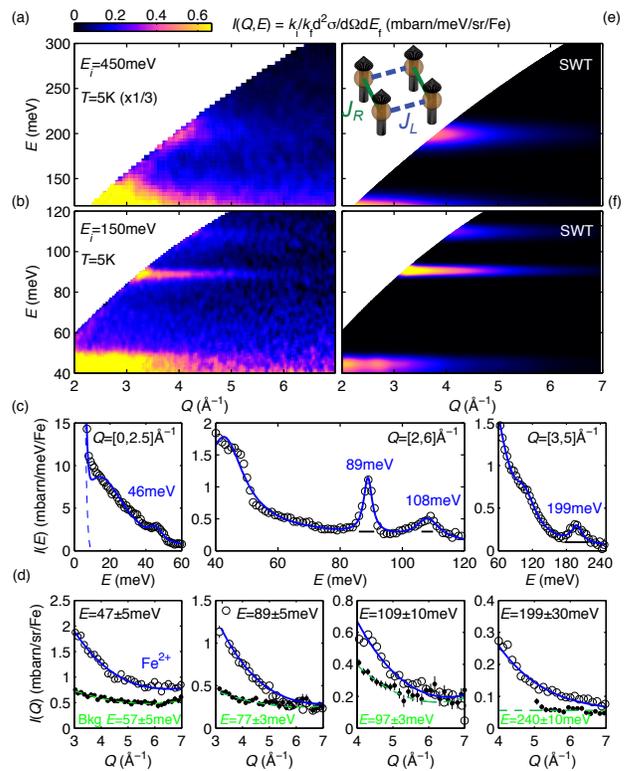}
\caption{(Color online) High energy spectrum of \bfs. 
(a-b) Intensity plot of $\tilde{I}(Q,E)$ at $T\!=\!5$~K with (a) $E_i\!=\!450$~meV and (b) 
$E_i\!=\!150$~meV. (c) Momentum-integrated inelastic scattering $\tilde{I}({\it E})$ (open symbols) for various ranges of $Q$ and fits to Lorentzian lineshapes (blue solid lines). (d) Energy-integrated inelastic scattering $\tilde{I}({\it Q})$ for the four modes (open symbols) compared to the nearby background $B(Q)$ (full symbols), and to the Fe$^{2+}$ form-factor $I(Q)= A |f(Q)|^2 + B(Q)$ (blue solid lines). (e-f) SWT prediction $\tilde{I}_{\rm SWT}(Q,E)$ for (e) $E_i\!=\!450$~meV and (f) $E_i\!=\!150$~meV in arbitrary units} \vspace{-0.6cm}
\label{fig3}
\end{figure}

To determine the bandwidth of the acoustic spin-waves in \bfs, we turn to the $E$-dependence of the low-energy signal $\tilde{I}({\it E})$ integrated over $0.6\!\leq\!Q\!\leq\!2.1$~\AA\ with $E_i\!=\!150$~meV [Fig.~\ref{fig2}(f)]. The low-energy excitations extend continuously from $E\approx\Delta$ up to $E\approx50$~meV, with a small peak at $E_{1}\!=\!46(1)$meV indicating the top of the acoustic spin wave band [see also Fig.~\ref{fig3}(c)]. As we shall see, this conventional 45~meV wide spectrum of acoustic spin waves belies the exotic Fe$_4$ block state. 

% --------------------------------------------------------------------
% --------------------------------------------------------------------

It is the higher-energy excitations of \bfs\ [Fig.~\ref{fig3}] that offer salient signatures of Fe$_4$ block magnetic order. With $E_i\!=\!450$~meV [Fig.~\ref{fig3}(a)] and $E_i\!=\!150$~meV [Fig.~\ref{fig3}(b)], the experiment covers a large dynamical range and reveals three additional bands of excitations, labeled $n\!=\!2,3$, and $4$ in the following. Two of these are centered around $E\!\approx\!100$~meV with $E_2\!=\!88.9(1)$~meV and $E_3\!=\!108.2(5)$~meV and the highest energy excitation is found at $E_4\!=\!198(1)$~meV [Fig.~\ref{fig3}(c)]. Their corresponding widths (Lorentzian FWHM), $\Gamma_n=$ 1.7(2) meV, 4(1) meV, and 15(3)~meV, respectively, can be compared with a geometry-based calculation of the energy-resolution of the spectrometer at $E\!=\!E_n$, $\delta E_n=$ 2.2(4) meV, 1.8(4) meV and 18(4)~meV, respectively. Although these over-estimate the resolution width by 20\%, they indicate the $n\!=\!2$ and $n\!=\!4$ modes are close to being resolution-limited while the $n\!=\!3$ excitation is intrinsically broad. The $Q$-dependence of all four bands of magnetic excitations [Fig.~\ref{fig3}({d})] compare well with the Fe$^{2+}$ form-factor for $Q\!\geq\!3$--4~\AA$^{-1}$, $\tilde{I}(Q)\!\propto\!|f(Q)|^2$ and contrasts with the approximatively $Q$-independent background. Given the form-factor and the fact that charge and intra-orbital $dd$-excitations (crystal-field excitations) have been observed by resonant inelastic X-Ray scattering (RIXS)~\cite{Monney2013a} at higher-energies, $E\!\approx\!0.35$~eV and $E\!\approx\!0.65$~eV respectively, we infer the signal arises from intra Fe$_4$ block excitations (optical spin waves).  

We can then extract the (total) dynamical spin correlation function for each band of scattering, $g^2\tilde{S}_n(Q,E)\!=\!6\tilde{I}_n(Q,E)/|r_0f(Q)|^2$ with $r_0\!=\!0.539 \times 10^{-12}$~cm, and subsequently obtain the inelastic spectral weight per Fe and per mode $\delta m_n^2\!=\mu_{\rm B}^2\!\int\!\int\!Q^2 [g^2\tilde{S}_n(Q,E)] {\rm d}Q{\rm d}E/\!\int\! Q^2{\rm d}Q $. After background subtraction and adapting the integration range to the bandwidth of each mode we obtain Tab.~\ref{tab1}. {Summing the observed static $\langle m \rangle^2$ and dynamic $\delta m^2$ spin correlations yields a total spectral weight of $m^2_{\rm tot}$=16(3)~$\mu_{\rm B}^2$ per Fe. This is significantly smaller than $g^2 S(S+1)\mu_{\rm B}^2\!=\!24$~$\mu_{\rm B}^2$ expected for $g\!=\!2$ and $S\!=\!2$. The total inelastic contribution of $8.2(2)$~$\mu_{\rm B}^2$ per Fe is however, larger than $g^2\langle S\rangle \mu_{\rm B}^2\!=\!5.2(4)\mu_{\rm B}^2$ which indicates an unusual magnetic ground-state in \bfs\ with a reduced ordered moment and enhanced spin fluctuations. Remarkably the total moment that we detect is consistent with the prediction of $16~\mu_{\rm B}^2$ per Fe obtained for the Fe$_4$ block in Ref.~\cite{Luo2013a} from a Hartree-Fock treatment of a five-band Hubbard model. 

% --------------------------------------------------------------------
\begin{table}[h!]
\begin{tabular}{|c|cccccc|} \hline
   Spectral Weight & $\langle m \rangle^2$ & $\delta m_1^2$  & $\delta m_2^2$ & $ \delta m_3^2$ 
   & $ \delta m_4^2$       & $m^2_{\rm tot}$  \\ \hline
   $\mu_{\rm B}^2.$ Fe$^{-1}$ & $7.5(3)$              & $5.8(1)$ & $0.5(1)$ & $0.9(1)$ 
   & $1.0(1)$ & $15.7(2)$   \\ \hline 
\end{tabular}
\caption{\label{tab1} Integrated elastic intensity $\langle m \rangle^2$ and inelastic spectral-weight $\delta m_n^2$ per spin-wave mode at $T\!=\!5$~K. The errorbars correspond to statistical uncertainty and do not include systematic errors in the normalization and background subtraction procedures estimated at 20\%.}\vspace{-0.4cm}
\end{table}
% --------------------------------------------------------------------

We now develop an effective spin-$S$ Heisenberg model for \bfs. We start from an isolated rectangular Fe$_4$ block with ferromagnetic $J_{\rm R}$ and $J_{\rm L}$ exchange interactions along its rungs and legs, respectively [Fig.~\ref{fig3}(e)(inset)]. An elementary diagonalization yields four localized excitations at energies $\tilde{\varepsilon}_{n}$ = 0, $2SJ_{\rm L}$, $2SJ_{\rm R}$ and $2S(J_{\rm R}+J_{\rm L})$. These resemble our observations of high-energy optical spin-waves with the ferromagnetic exchange parameters $SJ_{\rm L}\!\approx\!-44$~meV and $SJ_{\rm R}\!\approx\!-54$~meV or their permutation. The long range ordered state, the wide acoustic band, and the broadened 108~meV mode imply intra-ladder and inter-ladder exchange interactions that we parametrize consistent with the $Pnm2_1$ structure and the effects of which we describe with linear spin-wave theory (SWT) [Fig.~\ref{fig1}(b)]. These interactions can originate from Fe-Se-Fe and Fe-Se-Se-Fe super-exchange paths or from electronic ring-exchange terms that are indistinguishable from further-neighbor exchange at the level of linear SWT~\cite{DallaPiazza2012a}. 

Considering an isolated single-ladder with Fe$_4$ block spin structure, linear SWT yields four spin-wave modes ${\varepsilon}_{n}$ that directly stem from the above localized modes $\tilde{\varepsilon}_{n}$. Their energies are ${\varepsilon}_{n}(k)\!=\!S\sqrt{A_n^2 - B_n^2 \cos{(4\pi k)}}/\sqrt{2}$ where $A_n$ determines the average energy of each mode and $B_n\!=\!(\pm J_1^\prime \pm J_2^\prime \pm 2J_3)$  controls the bandwidth of their dispersion, with sign combinations $(+++)$, $(+-+)$, $(++-)$ and $(-++)$ for $n=1,2,3,$ and $4$, respectively. The constant $A_n$ depends on $J_{\rm R}, J_1, J_1^\prime, J_2, J_2^\prime$ and $J_3$. As for the isolated Fe$_4$ block, $J_{\rm R}$ and $J_1$ ($\equiv J_{\rm L}$) control the overall energy scale and the splitting between ${\varepsilon}_{2}$ and ${\varepsilon}_{3}$. As the lowest energy mode acquires a bandwidth controlled by $J_1^\prime+J_2^\prime+2J_3$, we expect reduced values for $J_{\rm R}$ and $J_1$ compared to an isolated plaquette. We also anticipate a sizable $J_2$ to account for the relative position of ${\varepsilon}_{2}$ and ${\varepsilon}_{3}$ with respect to ${\varepsilon}_{4}$. As $J_1^\prime$ and $J_2^\prime$ constrain the bandwidth of the high-energy modes, their values are important to allow a broad ${\varepsilon}_{3}$ while keeping the widths of ${\varepsilon}_{2}$ and ${\varepsilon}_{4}$ limited to the resolution of the instrument. 

To obtain realistic values for these exchanges, we compared the experimental energies $E_{n=1,2,3,4}$ and intrinsic widths $\Gamma_{n=2,3,4}$ to predictions from the analytical SWT model convoluted with the estimated instrumental resolution. A least-squares fit to the above {seven} experimental constraints yields $SJ_R\!=\!-43(2)$~meV, $SJ_1\!=\!-33(2)$~meV, $SJ_1^\prime\!=\!8(4)$~meV, $SJ_2\!=\!11(3)$~meV, $SJ_2^\prime\!=\!6(4)$~meV and $SJ_3\!=\!15(4)$ meV, see also Fig.~\ref{fig1}(b-c). Our model includes a small easy $\boldsymbol{a}$-axis anisotropy $|SD_a|\!=\!0.08(2)$~meV to account for the spin gap. In addition, the steep spin-wave dispersion and the absence of enhanced density-of-states (typically associated with inter-chain coupling) below 20~meV [Fig.~\ref{fig2}(a)], indicates inter-ladder exchanges greater than 3~meV. The latter is included in the above model through $SJ_5\!=\!4(1)$~meV but the data can be described without $J_4$ and $J_6$ [Fig.\ref{fig1}(a)]. Using the numerical implementation of linear SWT~\cite{Petit2011a,Toth2014a} in the \verb|SpinW| program~\cite{Toth2014a}, the powder averaged scattering intensity $\tilde{I}_{\rm SWT}(Q,E)$ for the model and exchanges of Fig.~\ref{fig1}(b-c) is shown in Fig.~\ref{fig2}(g) and Fig.~\ref{fig3}(e-f). The model accounts for all significant aspects of the data [Fig.~\ref{fig2}(a) and Fig.~\ref{fig3}(a-b)] {and in turn, the data provides evidence for all the magnetic excitations expected for the Fe$_4$ block state}. 

{We have shown that \bfs\ is a nearly spin-isotropic} {low}-dimensional antiferromagnet with a sizable ratio between inter-ladder and intra-ladder interactions $4J_5/(J_1^\prime+J_2^\prime+2J_3)\!\approx\!0.45$. We have determined a set of exchange interactions, compatible with the $Pnm2_1$ structure, that stabilizes the Fe$_4$ block ground-state and produces the peculiar multi-band excitation spectrum that we detected. Our experiment recovers a large fraction of, but not the entire, spectral weight expected for {localized Fe moments. Large missing neutron intensity was previously reported in insulating cuprates such as the spin-chain compound Sr$_2$CuO$_3$~\cite{Walters2009a} and attributed to hybridization between the magnetic Cu-$3d$ orbital and O-$2p$ orbitals. While our experiment cannot directly probe hybridization effects between the magnetic Fe-$3d$ and Se-$4p$ orbitals, the reduced effective moment observed in \bfs\ is remarkably consistent with the predictions for an orbital selective Mott state~\cite{Caron2012a,Rincon2014a}. This strongly favors a scenario where only $\approx\!2/3$ of the $3d$ electrons of Fe participate in the formation of local moments while the remaining electrons occupy wide electronic bands and remain beyond the energy range of the present experiment.} In this respect there are significant similarities to superconducting $\rm Rb_{0.89}Fe_{1.58}Se_{2}$~\cite{Wang2011a}. The sign reversal between effective intra-block ($J_1$) and inter-block ($J_1^\prime$) exchanges interactions is clear evidence for the orbital degrees of freedom that underlie a wealth of exotic magnetic and electronic ground-states in this class of materials. The microscopic spin Hamiltonian that we can report for \bfs, will advance a quantitative understanding of short range spin-orbital interactions in iron bearing square lattices and their potential role in promoting superconductivity.

%---------------------------------------------------------------------
% Acknowledgments
%---------------------------------------------------------------------
\begin{acknowledgments} 
The work at IQM was supported by the U.S. Department of Energy, Office of Basic Energy Sciences, Division of Material Sciences and Engineering under grant DE-FG02-08ER46544. This research at Oak Ridge National Laboratory's Spallation Neutron Source was sponsored by the U.S. Department of Energy, Office of Basic Energy Sciences, Scientific User Facilities Division. We are grateful to D. Abernathy for support on ARCS and to S. T\'oth for making \verb|SpinW| freely available.
\end{acknowledgments}
\vspace{-0.7cm}

%---------------------------------------------------------------------
% Bibliography from .bbl
%---------------------------------------------------------------------
%\bibliographystyle{prl}
%\bibliography{BaFe2Se3}
%

%---------------------------------------------------------------------
% End
%---------------------------------------------------------------------
\end{document}